\documentclass[10pt,doublecolumn]{IEEEtran}
\usepackage{amsfonts}
\usepackage{cite,graphicx,amsmath,amsthm}
\usepackage{subfigure}
\usepackage{fancyhdr}
\usepackage{dsfont}
\usepackage{array,color}
\usepackage{bm}
\usepackage{stfloats}

\newtheorem{lemma}{Lemma}

\newtheorem{proposition}{Proposition}

\newtheorem{corollary}{Corollary}

\newtheorem{property}{Property}

\newtheorem{remark}{Remark}

\newtheorem{claim}{Claim}

\begin{document}
\title{{Joint Relay-User Beamforming Design in Full-Duplex Two-Way Relay Channel}}

\author{Zhigang Wen,~Shuai Wang,~Xiaoqing Liu,~and Junwei Zou
\thanks{
Copyright (c) 2015 IEEE. Personal use of this material is permitted. However, permission to use this material for any other purposes must be obtained from the IEEE by sending a request to pubs-permissions@ieee.org.
This work was supported by the NSFC-61471067.

Zhigang Wen,~Xiaoqing Liu and Junwei Zou are with the Beijing Key Laboratory of Work Safety Intelligent Monitoring, School of Electronic Engineering, Beijing University of Posts and Telecommunications, Beijing 100876, P.R.China.

Shuai Wang (corresponding author) is with the Department of Electrical and Electronic Engineering,
The University of Hong Kong, Hong Kong (e-mail: swang@eee.hku.hk).

}
}

\markboth{}
{ \MakeLowercase{\textit{et al.}}: Joint Relay-User Beamforming Design in Full-Duplex Two-Way Relay Channel}

\maketitle

\vspace{-0.5in}

\begin{abstract}
A full-duplex two-way relay channel with multiple antennas is considered.
For this three-node network,
the beamforming design needs to suppress self-interference.
While a traditional way is to apply zero-forcing for self-interference mitigation, it may harm the desired signals.
In this paper, a design which \emph{reserves a fraction of self-interference} is proposed by solving a quality-of-service constrained beamforming design problem.
Since the problem is challenging due to the loop self-interference, a convergence-guaranteed alternating optimization algorithm is proposed to jointly design the relay-user beamformers.
Numerical results show that the proposed scheme outperforms zero-forcing method, and achieves a transmit power close to the ideal case.
\end{abstract}


\IEEEpeerreviewmaketitle

\section{Introduction}
\IEEEPARstart{F}ull-duplex (FD) is a promising technique to increase the spectral efficiency in relay systems.
However, the performance of FD relay suffers from the self interference (SI) \cite{1}.
Although natural isolation and time-domain cancellation can be applied for mitigation of SI, measurements show that residual SI still exists \cite{2}.
To this end, null-space projection using multiple antennas is proposed in one-way relay systems \cite{3}.
Furthermore, since zero-forcing (ZF) may harm the desired signal, relay beamforming design based on minimum mean square error (MMSE) is also proposed for one-way relays \cite{4}.

Very recently, FD two-way relay channel (TWRC) receives significant attentions \cite{5}, while the majority of literatures focus on the single antenna scenario \cite{6}.
To analyze the performance of multi-antenna FD TWRC, beamforming design based on ZF is proposed in \cite{7}.
However, it is known that the method in \cite{7} is suboptimal due to the manually added ZF constraint.
On the other hand, beamforming design which \emph{reserves a fraction of SI}
can overcome the drawback of \cite{7}, and currently has not been discussed in FD TWRC.
This paper provides the \emph{first attempt} to design such beamformers by solving the signal-to-interference-plus-noise ratio (SINR) quality-of-service (QoS) constrained beamforming design problem.
Compared with one-way relays, the considered problem in FD TWRC involves a single beamformer serving multiple users, and the MMSE based method in \cite{4} is not applicable.
Furthermore, due to the loop self-interference \cite{1}, the relay transmit power and users' SINRs are nonlinear fractional, leading to a challenging nonconvex problem.

To solve the nonconvex problem, a convergence guaranteed alternating optimization (AO) algorithm is proposed to divide it into four subproblems.
In particular, the subproblems of relay beamformers, user transmitters, and user receivers are solved using successive convex approximation (SCA), second-order cone programming (SOCP), and MMSE criterion, respectively.
Besides, a low-complexity ZF based design is presented as a benchmark.
Finally, numerical results show that the proposed algorithm outperforms existing algorithms, and achieves a transmit power close to the ideal case.

%The rest of the paper is organized as follows. In Section II, the system model is described and then the QoS problem is formulated.
%In Section III, the proposed AO algorithm is discussed in detail and proved to converge.
%To reduce the complexity, a low-complexity ZF-based scheme is also presented.
%Simulation results are presented in Section IV. Finally, conclusions are drawn in Section V.

\section{System Model and Problem Formulation}

\setcounter{secnumdepth}{4}We consider a FD-TWRC system consisting of a relay station and two users.
The relay, the first user, and the second user are equipped with $(M_R,M_1,M_2)$ transmit antennas and $(N_R,N_1,N_2)$ receive antennas, respectively.
As shown in Fig. 1, the two users have no direct link and wish to exchange information through the relay.
With the help of FD, the communication only requires one transmission phase, and all the nodes transmit their signals simultaneously.

\begin{figure}[!t]
\centering
\includegraphics[width=3in]{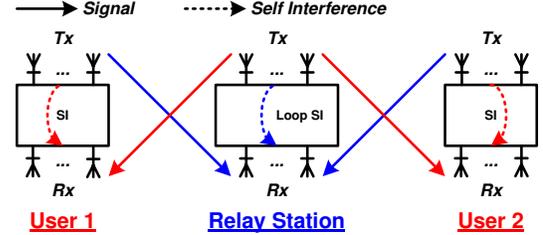}
% where an .eps filename suffix will be assumed under latex,
% and a .pdf suffix will be assumed for pdflatex; or what has been declared
% via \DeclareGraphicsExtensions.
\caption{System model of multi-antenna FD-TWRC.}
\label{fig_sim}
\end{figure}

We assume that the required processing time to implement the FD operation at relay is given by $\tau$-symbol duration, which is short compared to a time slot \cite{7}.
At the time instance $n$, the symbol $x_i[n]$ with $\mathbb{E}\big[|x_i[n]|\big]=1$ is transmitted at the $i^{\mathrm{th}}$ user, through the corresponding beamformer $\mathbf{f}_i\in\mathbb{C}^{M_i\times 1}$ with $||\mathbf{f}_i||^2=P_i$.
Meanwhile, the symbol $\mathbf{x}_R[n]$ is transmitted at the relay.
Therefore, the received signal $\mathbf{r}[n]\in\mathbb{C}^{N_R\times 1}$ at relay can be expressed as
\begin{eqnarray}\label{1}
\mathbf{r}[n]=\mathbf{H}_{1,R}\mathbf{f}_{1}x_{1}[n]+\mathbf{H}_{2,R}\mathbf{f}_{2}x_{2}[n]+\mathbf{H}_{R,R}\mathbf{x}_R[n]+\mathbf{n}[n],
\end{eqnarray}
where $\mathbf{H}_{i,R}\in\mathbb{C}^{N_R\times M_i}$ represents the channel from $i$ to R, and $\mathbf{H}_{R,R}\in\mathbb{C}^{N_R\times M_R}$ represents the residual SI channel.
The term $\mathbf{n}[n]\sim\mathcal{CN}(0,\sigma^2\mathbf{I})$ is the Gaussian noise at relay.
Since the received signal $\mathbf{r}[n-\tau]$ at time instant $n-\tau$ is filtered by receive beamformer $\mathbf{w}^H\in\mathbb{C}^{1\times N}$ and transmit beamformer $\mathbf{v}\in\mathbb{C}^{M_R\times 1}$, the transmitted signal at relay $\mathbf{x}_R[n]$ is
\begin{eqnarray}\label{2}
\mathbf{x}_R[n]=\mathbf{v}\mathbf{w}^H\mathbf{r}[n-\tau].
\end{eqnarray}
Putting equation \eqref{1} into \eqref{2}, we have
\begin{align}\label{3}
&\mathbf{x}_R[n]=
\mathbf{v}\sum_{r=0}^{\infty}(\mathbf{w}^H\mathbf{H}_{R,R}\mathbf{v})^r
\mathbf{w}^H\Big(
\mathbf{H}_{1,R}\mathbf{f}_{1}x_{1}[n-r\tau-\tau]
\nonumber\\
&~~~~~~~~~~
+
\mathbf{H}_{2,R}\mathbf{f}_{2}x_{2}[n-r\tau-\tau]
+\mathbf{n}[n-r\tau-\tau]
  \Big).
\end{align}

Observing that the power of $\mathbf{x}_R[n]$ in equation \eqref{3} is finite and $|\mathbf{w}^H\mathbf{H}_{R,R}\mathbf{v}|<1$,
we derive the relay output power as:
\begin{align}\label{Pr}
&P_r=\mathbb{E}(\mathbf{x}^H_R[n]\mathbf{x}_R[n])
\nonumber\\
&~~~
=\mathbf{v}^H\mathbf{v}
\mathop{\mathrm{lim}}_{n\rightarrow \infty}
\frac{1-|\mathbf{w}^H\mathbf{H}_{R,R}\mathbf{v}|^{2n}}{1-|\mathbf{w}^H\mathbf{H}_{R,R}\mathbf{v}|^2}
\Big(
|\mathbf{w}^H\mathbf{H}_{1,R}\mathbf{f}_{1}|^2
\nonumber\\
&~~~~~~~~~~~~~~~~~~~~~~~~~~~~~~~~~~~~
+|\mathbf{w}^H\mathbf{H}_{2,R}\mathbf{f}_{2}|^2
+\sigma^2\mathbf{w}^H\mathbf{w}
  \Big)
\nonumber\\
&
~~~=
\frac{\mathbf{v}^H\mathbf{v}
\Big(
|\mathbf{w}^H\mathbf{H}_{1,R}\mathbf{f}_{1}|^2+|\mathbf{w}^H\mathbf{H}_{2,R}\mathbf{f}_{2}|^2
+\sigma^2||\mathbf{w}||^2\Big)
}{1-|\mathbf{w}^H\mathbf{H}_{R,R}\mathbf{v}|^2}
,
\end{align}
where the equality in the third line is due to the sum formula of geometric sequence \cite{1}.
Based on \eqref{3}, the received signal $\mathbf{y}_i[n]$ at the $i^{\mathrm{th}}$ user in the downlink phase is
\begin{align}
&\mathbf{y}_i[n]=\mathbf{H}_{R,i}\mathbf{x}_R[n]+\mathbf{H}_{i,i}\mathbf{f}_{i}x_{i}[n]+\mathbf{z}_i[n]
\nonumber
\\
&=\mathbf{H}_{R,i}\mathbf{v}
\Big(\mathbf{w}^H\mathbf{H}_{1,R}\mathbf{f}_{1}x_{1}[n-\tau]+\mathbf{w}^H\mathbf{H}_{2,R}\mathbf{f}_{2}x_{2}[n-\tau]
\Big)
\nonumber\\
&
+
\mathbf{H}_{R,i}\mathbf{v}\sum_{r=1}^{\infty}(\mathbf{w}^H\mathbf{H}_{R,R}\mathbf{v})^r
\mathbf{w}^H\Big(
\mathbf{H}_{1,R}\mathbf{f}_{1}x_{1}[n-r\tau-\tau]
\nonumber\\
&~~~~~~~~~~~~~~~~~~~~~~~~~~~~~~~~~~~~~~
+\mathbf{H}_{2,R}\mathbf{f}_{2}x_{2}[n-r\tau-\tau]
\Big)
\nonumber\\
&
+\mathbf{H}_{R,i}\mathbf{v}
\sum_{r=0}^{\infty}(\mathbf{w}^H\mathbf{H}_{R,R}\mathbf{v})^r
\mathbf{w}^H\mathbf{n}[n-r\tau-\tau]
\nonumber\\
&
+\mathbf{H}_{i,i}\mathbf{f}_{i}x_{i}[n]+\mathbf{z}_i[n], \label{yi}
\end{align}
where $\mathbf{H}_{R,i}\in\mathbb{C}^{N_i\times M_R}$ represents the channel from relay to user $i$, and $\mathbf{H}_{i,i}\in\mathbb{C}^{N_i\times M_i}$ represents the residual SI channel at user $i$.
The term $\mathbf{z}_i[n]\sim\mathcal{CN}(0,\sigma^2\mathbf{I})$ is Gaussian noise at the $i^{\mathrm{th}}$ user.
Since the first term in \eqref{yi} is the desired signal and the residual is the interference, by applying user receiver $\mathbf{u}^H_i\in\mathbb{C}^{1\times N_i}$ with $||\mathbf{u}_i||=1$ to $\mathbf{y}_i[n]$ in \eqref{yi}, the SINR $\gamma_i$ at the $i^{\mathrm{th}}$ user can be expressed as equation (6).
\begin{figure*}[ht]
\begin{align}\label{gamma}
&\gamma_i(\mathbf{v},\mathbf{w},\{\mathbf{f}_i,\mathbf{u}_i\})
=
\dfrac{|\mathbf{u}^H_i\mathbf{H}_{R,i}\mathbf{v}|^2
|\mathbf{w}^H\mathbf{H}_{3-i,R}\mathbf{f}_{3-i}|^2}
{\dfrac{
|\mathbf{u}^H_i\mathbf{H}_{R,i}\mathbf{v}|^2|\mathbf{w}^H\mathbf{H}_{R,R}\mathbf{v}|^2}
{1-|\mathbf{w}^H\mathbf{H}_{R,R}\mathbf{v}|^2}
(|\mathbf{w}^H\mathbf{H}_{1,R}\mathbf{f}_{1}|^2+|\mathbf{w}^H\mathbf{H}_{2,R}\mathbf{f}_{2}|^2)
+\sigma^2
\dfrac{|\mathbf{u}^H_i\mathbf{H}_{R,i}\mathbf{v}|^2||\mathbf{w}||^2}{1-|\mathbf{w}^H\mathbf{H}_{R,R}\mathbf{v}|^2}+
|\mathbf{u}^H_i\mathbf{H}_{i,i}\mathbf{f}_{i}|^2
+\sigma^2}.
\end{align}
\hrulefill
\end{figure*}

To provide reliable communication for both users at their required SINR threshold, we must have $\gamma_i\geq\theta_i$, where $\theta_i$ is the SINR requirement at the $i^{\mathrm{th}}$ user.
On the other hand, having the QoS requirement satisfied, it is crucial to reduce the total transmit power for cost reduction and environment benefits.
Therefore, we can write the SINR QoS constrained beamforming design problem as $\mathcal{P}1$.
The optimization problem $\mathcal{P}1$ is carried out at the relay node.
The channel state information (CSI) of $\mathbf{H}_{i,R},\mathbf{H}_{R,R}$ is available at relay using pilots.
The CSI of $\mathbf{H}_{R,i}$ can be obtained with channel reciprocity.
The CSI of $\mathbf{H}_{i,i}$ can be obtained from feedback of users.
However, $\mathcal{P}1$ is difficult to solve because variables $\mathbf{v},\mathbf{w}$, $\mathbf{f}_i$ and $\mathbf{u}_i$ are coupled, and the terms of $\mathbf{v},\mathbf{w}$ are nonlinear fractional.
To decouple the variables $\mathbf{v}$, $\mathbf{w}$, $\{\mathbf{f}_i\}$ and $\{\mathbf{u}_i\}$, we propose the AO algorithm \cite{8} in the next section.
\begin{figure*}[ht]
\begin{align}
&~~\mathcal{P}1:\mathop{\mathrm{min}}_{\substack{\mathbf{v},\mathbf{w},\{\mathbf{f}_i,\mathbf{u}_i\}}}
~
\frac{\mathbf{v}^H\mathbf{v}}{1-|\mathbf{w}^H\mathbf{H}_{R,R}\mathbf{v}|^2}
\Big(
|\mathbf{w}^H\mathbf{H}_{1,R}\mathbf{f}_{1}|^2+|\mathbf{w}^H\mathbf{H}_{2,R}\mathbf{f}_{2}|^2
+\sigma^2||\mathbf{w}||^2\Big)+\mathbf{f}^H_{1}\mathbf{f}_{1}+\mathbf{f}^H_{2}\mathbf{f}_{2}
\nonumber\\
&~~~~~~~~~~~~~\mathrm{s.t.}~~~~
\gamma_i(\mathbf{v},\mathbf{w},\{\mathbf{f}_i,\mathbf{u}_i\})
\geq\theta_i,~\forall i=1,2,~|\mathbf{w}^H\mathbf{H}_{R,R}\mathbf{v}|^2<1,~||\mathbf{u}_i||=1. \nonumber
\end{align}
\hrulefill
\end{figure*}

\section{Joint Relay-User Beamformer With Alternating Optimization }

\subsection{Subproblem of $\mathbf{v}$}
This subsection discusses the optimization of $\mathbf{v}$ when other variables are fixed.
To simplify the notation, let $g_{i,R}=\mathbf{w}^H\mathbf{H}_{i,R}\mathbf{f}_i$, $g_{i,i}=|\mathbf{u}_{i}\mathbf{H}^H_{i,i}\mathbf{f}_{i}|^2$, $\mathbf{g}^H_{R,i}=\mathbf{u}_i^H\mathbf{H}_{R,i}$, and $\mathbf{g}^H_{R,R}=\mathbf{w}^H\mathbf{H}_{R,R}$.
Then problem $\mathcal{P}1$ reduces to
\begin{subequations}
\begin{align}
&\mathcal{P}2:\mathop{\mathrm{min}}_{\substack{\mathbf{v}}}
~\Big(g_{1,R}+g_{2,R}
+\sigma^2\Big)
\frac{\mathbf{v}^H\mathbf{v}}{1-|\mathbf{g}^H_{R,R}\mathbf{v}|^2}
\label{P1.1-a}\\
&\mathrm{s.t.}~~g_{3-i,R}|\mathbf{g}^H_{R,i}\mathbf{v}|^2
+\theta_i(g_{i,i}+\sigma^2)|\mathbf{g}^H_{R,R}\mathbf{v}|^2
\geq
\nonumber\\
&~~~~~~~~~~~
\Big(\theta_i(g_{1,R}+g_{2,R})+g_{3-i,R}\Big)
|\mathbf{g}^H_{R,R}\mathbf{v}|^2|\mathbf{g}^H_{R,i}\mathbf{v}|^2
\nonumber\\
&~~~~~~~~~~~
+\theta_i
\sigma^2||\mathbf{w}||^2|\mathbf{g}^H_{R,i}\mathbf{v}|^2
+\theta_i(g_{i,i}+\sigma^2),~\forall i\label{P1.1-b}
\\
&~~~~~~
|\mathbf{g}^H_{R,R}\mathbf{v}|^2<1.
\end{align}
\end{subequations}

Problem $\mathcal{P}2$ is challenging because the objective function is quadratic fractional and the constraints are quartic.
To tackle it, we introduce a slack variable $\xi\geq\mathbf{v}^H\mathbf{v}/(1-|\mathbf{g}^H_{R,R}\mathbf{v}|^2)$,
which is equivalent to
\begin{equation}\label{a-1}
\left[
\begin{array}{cccc}
1-|\mathbf{g}^H_{R,R}\mathbf{v}|^2 & \mathbf{v}^H
\\
\mathbf{v} & \xi\mathbf{I}
\end{array}
\right]
\succeq 0
\end{equation}
according to Schur Complement Lemma \cite{9}.
Then the objective function \eqref{P1.1-a}
becomes $\Big(g_{1,R}+g_{2,R}+\sigma^2\Big)\xi$, which is linear in $\xi$.
To further transform \eqref{a-1} into a linear matrix inequality (LMI), we can introduce another slack variable $\mu$ such that
$
1-|\mathbf{g}^H_{R,R}\mathbf{v}|^2\geq \mu,
$
and this inequality is convex.

On the other hand, we deal with the constraint \eqref{P1.1-b} as follows.
Specifically, rearrange the left hand side of \eqref{P1.1-b} as
\begin{align}\label{14}
&g_{3-i,R}|\mathbf{g}^H_{R,i}\mathbf{v}|^2
+\theta_i(g_{i,i}+\sigma^2)|\mathbf{g}^H_{R,R}\mathbf{v}|^2
\nonumber\\
&
=
\mathbf{v}^H\underbrace{\Big(g_{3-i,R}\mathbf{g}_{R,i}\mathbf{g}^H_{R,i}+\theta_i(g_{i,i}+\sigma^2)\mathbf{g}_{R,R}\mathbf{g}^H_{R,R}\Big)
}_{\mathbf{\Phi}_i}\mathbf{v}.
\end{align}
Moreover, to deal with the quartic term on the right hand side of \eqref{P1.1-b}, introduce
slack variable $\lambda_i\geq|\mathbf{g}^H_{R,R}\mathbf{v}|^2|\mathbf{g}^H_{R,i}\mathbf{v}|^2$.
Since $\mathbf{g}^H_{R,i}\mathbf{v}\neq 0$, this newly added inequality can be recast as an LMI
\begin{equation}
\left[
\begin{array}{cccc}
\lambda_i & \mathbf{g}^H_{R,R}\mathbf{v}
\\
\mathbf{v}^H\mathbf{g}_{R,R} & \rho_i
\end{array}
\right]
\succeq 0,
\end{equation}
where $\rho_i$ is a slack variable with $|\mathbf{g}^H_{R,i}\mathbf{v}|^{-2}\geq\rho_i$.

With the above procedure, problem $\mathcal{P}2$ is equivalently transformed into
\begin{subequations}
\begin{align}
&\mathcal{P}2':\mathop{\mathrm{min}}_{\substack{\mathbf{v},\xi,\mu,\{\rho_i,\lambda_i\}}}
~\xi
\\
&\mathrm{s.t.}~~\mathbf{v}^H\mathbf{\Phi}_i\mathbf{v}
\geq
\Big(\theta_i(g_{1,R}+g_{2,R})+g_{3-i,R}\Big)\lambda_i
\nonumber\\
&~~~~~~~~~~~~~
+
\theta_i
\sigma^2||\mathbf{w}||^2|\mathbf{g}^H_{R,i}\mathbf{v}|^2
+\theta_i(g_{i,i}+\sigma^2),~~\forall i
\label{concave1}
\\
&~~~~~~
\frac{1}{\rho_i}\geq|\mathbf{g}^H_{R,i}\mathbf{v}|^2,~
\left[
\begin{array}{cccc}
\lambda_i & \mathbf{g}^H_{R,R}\mathbf{v}
\\
\mathbf{v}^H\mathbf{g}_{R,R} & \rho_i
\end{array}
\right]
\succeq 0,
~~\forall i
\label{concave}
\\
&~~~~~
\left[
\begin{array}{cccc}
\mu & \mathbf{v}^H
\\
\mathbf{v} & \xi\mathbf{I}
\end{array}
\right]
\succeq 0,~\mu+|\mathbf{g}^H_{R,R}\mathbf{v}|^2\leq1
.\label{2.1d}
\end{align}
\end{subequations}
Problem $\mathcal{P}2'$ is still nonconvex due to term $\mathbf{v}^H\mathbf{\Phi}_i\mathbf{v}$ in \eqref{concave1} and $1/\rho_i$ in \eqref{concave}.
However, we can apply SCA to construct a linear approximation for them.
In particular, define $\Upsilon_{i}(\mathbf{v}):=
\mathbf{v}^H\mathbf{\Phi}_i\mathbf{v}$, and
$\Delta_{i}(\rho_i):=1/\rho_i$.
Assuming that the solution at the $n^{\mathrm{th}}$ iteration is given by $\mathbf{v}^{[n]},\{\rho^{[n]}_{i}\}$\footnote{
The superscript $[n]$ in this subsection refers to the index of inner SCA iteration, and the initial $\mathbf{v}^{[0]},\{\rho^{[0]}_{i}\}$ is obtained from last iteration of AO.
},
now define functions
\begin{align}\label{gapp}
&\tilde{\Upsilon}^{[n]}_{i}(\mathbf{v})=2\mathrm{Re}[(\mathbf{v}^{[n]})^H\mathbf{\Phi}_i\mathbf{v}]-(\mathbf{v}^{[n]})^H\mathbf{\Phi}_i\mathbf{v}^{[n]},
\nonumber\\
&\tilde{\Delta}^{[n]}_{i}(\rho_i)=\frac{2}{\rho_i^{[n]}}-\frac{1}{(\rho_i^{[n]})^2}\rho_i,
\end{align}
and the following property can be established.
\begin{property}
The functions $\tilde{\Upsilon}^{[n]}_{i}$ and $\tilde{\Delta}^{[n]}_{i}$ satisfy the following:
(i) $\tilde{\Upsilon}^{[n]}_{i}(\mathbf{v})\leq \Upsilon_{i}(\mathbf{v}),\tilde{\Delta}^{[n]}_{i}(\rho_i)\leq \Delta_{i}(\rho_i)$;
(ii) $\tilde{\Upsilon}^{[n]}_{i}(\mathbf{v}^{[n]})=\Upsilon_{i}(\mathbf{v}^{[n]}), \tilde{\Delta}^{[n]}_{i}(\rho^{[n]}_i)=\Delta_{i}(\rho^{[n]}_i)$; and
(iii)
\begin{align}
&\dfrac{\partial \tilde{\Upsilon}^{[n]}_{i}(\mathbf{v})}{\partial\mathbf{v}}
\Big|_{\mathbf{v}=\mathbf{v}^{[n]}}
=
\dfrac{\partial \Upsilon_{i}(\mathbf{v})}{\partial\mathbf{v}}
\Big|_{\mathbf{v}=\mathbf{v}^{[n]}}
,
\nonumber\\
&
\dfrac{\partial \tilde{\Delta}^{[n]}_{i}(\rho_i)}{\partial\rho_{i}}
\Big|_{\rho_{i}=\rho^{[n]}_{i}}
=
\dfrac{\partial \Delta_{i}(\rho_{i})}{\partial\rho_{i}}
\Big|_{\rho_{i}=\rho^{[n]}_{i}}.\nonumber
\end{align}
\end{property}
\begin{proof}
See Appendix A.
\end{proof}
With the result of \textbf{Property 1}, the following problem is considered at the $(n+1)^{\mathrm{th}}$ iteration:
\begin{align}
&~~~~~~\mathcal{P}2'[n+1]:\mathop{\mathrm{min}}_{\substack{\mathbf{v},\xi,\mu,\{\rho_i,\lambda_i\}}}
~\xi\nonumber
\\
&\mathrm{s.t.}~\tilde{\Upsilon}^{[n]}_{i}
\geq
\Big(\theta_i(g_{1,R}+g_{2,R})+g_{3-i,R}\Big)\lambda_i
\nonumber\\
&~~~~~~~~~~~~~~
+\theta_i
\sigma^2||\mathbf{w}||^2|\mathbf{g}^H_{R,i}\mathbf{v}|^2
+\theta_i(g_{i,i}+\sigma^2),~~\forall i
\nonumber
\\
&~~~~~
\tilde{\Delta}^{[n]}_{i}\geq|\mathbf{g}^H_{R,i}\mathbf{v}|^2,~
\left[
\begin{array}{cccc}
\lambda_i & \mathbf{g}^H_{R,R}\mathbf{v}
\\
\mathbf{v}^H\mathbf{g}_{R,R} & \rho_i
\end{array}
\right]
\succeq 0,
~~\forall i,~\eqref{2.1d}.
\nonumber
\end{align}
Problem $\mathcal{P}2'[n+1]$ is a semidefinite programming (SDP), which can be optimally solved by CVX, a Matlab-based software for convex optimization \cite{12}.
Denoting the solution of $\mathcal{P}2'[n+1]$ as $\mathbf{v}^*,\xi^*,\mu^*,\{\rho_i^*,\lambda^*_i\}$ and setting $\mathbf{v}^{[n+1]}=\mathbf{v}^*$ and $\rho^{[n+1]}_i=\rho^*_i$, we can proceed to solve $\mathcal{P}2'[n+2]$.
According to \textbf{Property 1} and \cite[Theorem 1]{10}, the iterative algorithm is convergent, and the converged point would be a local optimal solution for $\mathcal{P}2$ based on \cite{boyd}.

\subsection{Subproblem of $\mathbf{w}$}
This subsection discusses the optimization of $\mathbf{w}$ when other variables are fixed.
Let
$\mathbf{q}_{i,R}=\mathbf{H}_{i,R}\mathbf{f}_i$,
$q_{i,i}=|\mathbf{u}_i^H\mathbf{H}_{i,i}\mathbf{f}_i|^2$,
$q_{R,i}=|\mathbf{u}_i^H\mathbf{H}_{R,i}\mathbf{v}|^2$,
and
$\mathbf{q}_{R,R}=\mathbf{H}_{R,R}\mathbf{v}$.
Then problem $\mathcal{P}1$ reduces to
\begin{subequations}
\begin{align}
&\mathcal{P}3:\mathop{\mathrm{min}}_{\substack{\mathbf{w}}}
~||\mathbf{v}||^2\cdot
\frac{|\mathbf{w}^H\mathbf{q}_{1,R}|^2+|\mathbf{w}^H\mathbf{q}_{2,R}|^2+\sigma^2\mathbf{w}^H\mathbf{w}}
{1-|\mathbf{w}^H\mathbf{q}_{R,R}|^2} \nonumber
\\
&\mathrm{s.t.}~~
q_{R,i}|\mathbf{w}^H\mathbf{q}_{3-i,R}|^2
\geq
\nonumber\\
&~~~~~~
q_{R,i}|\mathbf{w}^H\mathbf{q}_{R,R}|^2\Big(|\mathbf{w}^H\mathbf{q}_{3-i,R}|^2
+\theta_i\sum_{j=1}^2|\mathbf{w}^H\mathbf{q}_{j,R}|^2\Big)
\nonumber\\
&~~~~~
+\theta_i\sigma^2q_{R,i}||\mathbf{w}||^2
+\theta_i(q_{i,i}+\sigma^2)(1-|\mathbf{w}^H\mathbf{q}_{R,R}|^2),~\forall i \nonumber
\\
&~~~~~~
|\mathbf{w}^H\mathbf{q}_{R,R}|^2<1. \nonumber
\end{align}
\end{subequations}
Problem $\mathcal{P}3$ has the same structure with $\mathcal{P}2$, and we can find the local optimal solution of $\mathbf{w}$.
Due to space limitation, the details are omitted here.

\subsection{Subproblem of $\mathbf{f}_i$}
Now suppose that $\mathbf{v},\mathbf{w}$ and $\mathbf{u}_i$ are given.
Let
$\mathbf{a}^H_{i,R}=\mathbf{w}^H\mathbf{H}_{i,R}$,
$\mathbf{a}^H_{i,i}=\mathbf{u}_i^H\mathbf{H}_{i,i}$,
$a_{R,i}=|\mathbf{u}_i^H\mathbf{H}_{R,i}\mathbf{v}|^2$,
and
$a_{R,R}=|\mathbf{w}^H\mathbf{H}_{R,R}\mathbf{v}|^2$.
Then problem $\mathcal{P}1$ reduces to
\begin{subequations}
\begin{align}
&\mathcal{P}4:\mathop{\mathrm{min}}_{\substack{\{\mathbf{f}_i\}}}
~
\frac{||\mathbf{v}||^2}{1-a_{R,R}}
\Big(
\sum_{j=1}^2
|\mathbf{a}^H_{j,R}\mathbf{f}_{j}|^2+\sigma^2\Big)
+\sum_{j=1}^2||\mathbf{f}_{j}||^2
\nonumber
\\
&\mathrm{s.t.}~~(1-a_{R,R})a_{R,i}
|\mathbf{a}^H_{3-i,R}\mathbf{f}_{3-i}|^2\geq
\nonumber\\
&~~~~~~~~~~~~~
\theta_i
a_{R,i}a_{R,R}
(|\mathbf{a}^H_{1,R}\mathbf{f}_{1}|^2+|\mathbf{a}^H_{2,R}\mathbf{f}_{2}|^2)
\nonumber\\
&~~~~~~~~~~~~
+\theta_i(1-a_{R,R})(|\mathbf{a}^H_{i,i}\mathbf{f}_{i}|^2+\sigma^2)+\theta_i\sigma^2a_{R,i},~~\forall i.
\nonumber
\end{align}
\end{subequations}
The objective function of $\mathcal{P}4$ is convex quadratic with respect to $\mathbf{f}_i$.
On the other hand,
by taking the square root of the constraints of $\mathcal{P}4$ on both sides and replacing $
|\mathbf{a}^H_{3-i,R}\mathbf{f}_{3-i}|$ with $
\mathrm{Re}(\mathbf{a}^H_{3-i,R}\mathbf{f}_{3-i})$ according to phase rotation \cite{13},
the constraints of $\mathcal{P}4$ can be equivalently reformulated as
\begin{align}\label{15}
&\sqrt{(1-a_{R,R})a_{R,i}}\cdot
\mathrm{Re}(\mathbf{a}^H_{3-i,R}\mathbf{f}_{3-i})\geq
\Big(|\sqrt{\theta_i
a_{R,i}a_{R,R}}\mathbf{a}^H_{1,R}\mathbf{f}_{1}|^2
\nonumber\\
&~~~+
|\sqrt{\theta_i
a_{R,i}a_{R,R}}\mathbf{a}^H_{2,R}\mathbf{f}_{2}|^2
+|\sqrt{\theta_i(1-a_{R,R})}\mathbf{a}^H_{i,i}\mathbf{f}_{i}|^2
\nonumber\\
&~~~
+
|\sqrt{\theta_i(1-a_{R,R})\sigma^2+\theta_i a_{R,i}\sigma^2}|^2\Big)^{1/2},
~~\forall i.
\end{align}
With constraint \eqref{15}, problem $\mathcal{P}4$ can be reformulated as an SOCP and solved by CVX \cite{12}.

\subsection{Subproblem of $\mathbf{u}_i$}
Suppose that $\mathbf{v},\mathbf{w}$ and $\mathbf{f}_i$ are given.
Since $\mathbf{u}_i$ is only involved in $\gamma_i$,
the optimal $\mathbf{u}_i$ can be derived by maximizing $\gamma_i$ in \eqref{gamma}.
Specifically, dividing $|\mathbf{u}^H_i\mathbf{H}_{R,i}\mathbf{v}|^2$ by the numerator and denominator of \eqref{gamma}, the optimal $\mathbf{u}_i$ is the MMSE receiver \cite{14} with the following closed form:
\begin{align}
&\mathbf{u}^*_i=\sqrt{\beta_i}\Big(\sigma^2\mathbf{I}_{N_i}+\mathbf{H}_{i,i}\mathbf{f}_i\mathbf{f}^H_i\mathbf{H}^H_{i,i}
  \Big)^{-1}
  \mathbf{H}_{R,i}\mathbf{v},
\end{align}
where $\mathbf{I}_{N_i}$ is an $N_i\times N_i$ identity matrix, $\beta_i$ is a normalization coefficient such that $||\mathbf{u}^*_i||=1$.

\subsection{Convergence and Complexity Analysis}
Based on the solutions for subproblems, we develop the following iterative algorithm for problem $\mathcal{P}1$.
That is, initializing the variables as $\mathbf{v}^{[0]},\mathbf{w}^{[0]},\{\mathbf{f}^{[0]}_i,\mathbf{u}^{[0]}_i\}$ based on Appendix B, and then optimizing $\mathbf{v},\mathbf{w},\mathbf{f}_i,\mathbf{u}_i$ alternatively according to Section A-D.
Since the proposed method is an inexact AO algorithm (subproblems of $\mathcal{P}2$ and $\mathcal{P}3$ are not solved optimally), the convergence is not straightforward.
To this end, we establish the following property.

\begin{property}
The proposed AO algorithm is convergent.
\end{property}
\begin{proof}
At the $m^{\mathrm{th}}$ iteration,
given $\mathbf{w}^{[m-1]},\{\mathbf{f}^{[m-1]}_i\},$ and $\{\mathbf{u}^{[m-1]}_i\}$, we update $\mathbf{v}^{[m]}$ using SCA.
Since the initial point is $\mathbf{v}^{[m-1]},\{\rho^{[m-1]}_i=|\mathbf{h}_{R,i}\mathbf{v}^{[m-1]}|^{-2}\}$, and SCA is guaranteed to produce a monotonically decreasing sequence \cite{10}, the objective value corresponding to $\mathbf{v}^{[m]},\mathbf{w}^{[m-1]},\{\mathbf{f}^{[m-1]}_i,\mathbf{u}^{[m-1]}_i\}$ is no larger than that of
$\mathbf{v}^{[m-1]},\mathbf{w}^{[m-1]},\{\mathbf{f}^{[m-1]}_i,\mathbf{u}^{[m-1]}_i\}$.
After we obtain $\mathbf{v}^{[m]}$, update $\mathbf{w}^{[m]}$ by solving problem $\mathcal{P}3$ with SCA.
Similarly, the objective value corresponding to $\mathbf{v}^{[m]},\mathbf{w}^{[m]},\{\mathbf{f}^{[m-1]}_i,\mathbf{u}^{[m-1]}_i\}$ would be no larger than that of
$\mathbf{v}^{[m]},\mathbf{w}^{[m-1]},\{\mathbf{f}^{[m-1]}_i,\mathbf{u}^{[m-1]}_i\}$.
Finally, we update $\mathbf{f}_i^{[m]}$ and $\mathbf{u}_i^{[m]}$ based on SCOP in \emph{Section C} and MMSE in \emph{Section D}.
Since the two subproblems are optimally solved, the objective value corresponding to $\mathbf{v}^{[m]},\mathbf{w}^{[m]},\{\mathbf{f}^{[m]}_i,\mathbf{u}^{[m]}_i\}$ would be no larger than that of $\mathbf{v}^{[m]},\mathbf{w}^{[m]},\{\mathbf{f}^{[m-1]}_i,\mathbf{u}^{[m-1]}_i\}$.
Therefore, the whole AO procedure will produce a monotonically decreasing sequence of the objective values, and the sequence is lower bounded by zero, which proves the convergence of the AO algorithm.
\end{proof}

In terms of complexity, the AO algorithm is dominated by the subproblems of $\mathbf{v}$ and $\mathbf{w}$.
Specifically, solving the problem $\mathcal{P}2'[n+1]$ requires complexity $O((M_R+1)^{3.5})$ \cite{12}.
For $p_1$ iterations of SCA, the complexity of solving $\mathcal{P}2$ would be $O(p_1(M_R+1)^{3.5})$,
and that of solving $\mathcal{P}3$ would be $O(p_1(N_R+1)^{3.5})$.
Therefore, the total complexity is $O\Big(p_2[p_1(M_R+1)^{3.5}+p_1(N_R+1)^{3.5}]\Big)$, where $p_2$ is the number of iterations for AO.

\subsection{Low-Complexity Zero-Forcing Design}
This subsection presents a low-complexity ZF based algorithm as a benchmark, which is a generalized version of \cite{7}.
Specifically, the users apply $\mathbf{u}_i^H\mathbf{H}^H_{i,i}\mathbf{f}_i=0$ to cancel the SI
,
and the relay applies $\mathbf{w}^H\mathbf{H}^H_{R,R}\mathbf{v}=0$ to cancel the loop SI.
Therefore the problem $\mathcal{P}1$ reduces to
\begin{subequations}
\begin{align}
&~~\mathcal{P}5: \mathop{\mathrm{min}}_{\substack{\mathbf{v},\mathbf{w},\{\mathbf{f}_i,\mathbf{u}_i\}}}
~
\mathbf{v}^H\mathbf{v}
\Big(
|\mathbf{w}^H\mathbf{H}_{1,R}\mathbf{f}_{1}|^2+|\mathbf{w}^H\mathbf{H}_{2,R}\mathbf{f}_{2}|^2
\nonumber\\
&~~~~~~~~~~~~~~~~~~~~~~~~~~~~~
+\sigma^2||\mathbf{w}||^2\Big)+\mathbf{f}^H_{1}\mathbf{f}_{1}+\mathbf{f}^H_{2}\mathbf{f}_{2}
\nonumber
\\
&~~~\mathrm{s.t.}~~
|\mathbf{u}^H_i\mathbf{H}_{R,i}\mathbf{v}|^2
|\mathbf{w}^H\mathbf{H}_{3-i,R}\mathbf{f}_{3-i}|^2
\geq
\nonumber\\
&~~~~~~~~~
\theta_i\Big(\sigma^2|\mathbf{u}^H_i\mathbf{H}_{R,i}\mathbf{v}|^2||\mathbf{w}||^2+\sigma^2\Big),~\forall i,
\nonumber\\
&~~~~~~~~~
\mathbf{u}_i^H\mathbf{H}_{i,i}\mathbf{f}_i=0,~||\mathbf{u}_i||=1,~~\forall i,~
\mathbf{w}^H\mathbf{H}_{R,R}\mathbf{v}=0.
\nonumber
\end{align}
\end{subequations}
Following the aforesaid AO algorithm, $\mathcal{P}5$ can be divided into four subproblems as well, and all the subproblems have closed-form optimal solutions.
Specifically, the closed-form solutions of $\mathbf{w}$ and $\mathbf{v}$ are given in \cite[Section III-A]{7} and \cite[Section III-B]{7}, respectively.

To derive the closed-form solution of $\{\mathbf{f}_i\}$,
applying singular value decomposition (SVD) to $\mathbf{H}^H_{i,i}\mathbf{u}_i$, we have
$\mathbf{H}^H_{i,i}\mathbf{u}_i=[\mathbf{z}_{i}~\mathbf{Z}_{i}]\mathbf{\Lambda}_i\mathbf{V}_{i}^H$,
where $\mathbf{Z}_{i}\in\mathbb{C}^{M_i\times (M_i-1)}$.
Using the null space matrix $\mathbf{Z}_{i}$,
we perform SVD
$\mathbf{w}^H\mathbf{H}_{i,R}\mathbf{Z}_{i}=\mathbf{A}_{i}\mathbf{B}_i[\mathbf{c}_{i}~\mathbf{C}_{i}]^H$,
where $\mathbf{c}_{i}\in\mathbb{C}^{(M_i-1)\times 1}$.
Then, the optimal ZF user beamformer is
$\mathbf{f}^*_{i}=\sqrt{\alpha_{i}}\mathbf{Z}_{i}\mathbf{c}_{i}
$, where
\begin{align}
\alpha_{i}=
\frac{\theta_{3-i}\Big(\sigma^2|\mathbf{u}^H_{3-i}\mathbf{H}^H_{R,3-i}\mathbf{v}|^2||\mathbf{w}||^2+\sigma^2\Big)}
{|\mathbf{u}^H_{3-i}\mathbf{H}_{R,3-i}\mathbf{v}|^2|\mathbf{w}^H\mathbf{H}_{i,R}\mathbf{Z}_{i}
\mathbf{c}_{i}
|^2}.\nonumber
\end{align}
The closed-form solution of $\{\mathbf{u}_i\}$ can be similarly derived using the null space of $\mathbf{H}_{i,i}\mathbf{f}_i$
and range space of $\mathbf{H}_{R,i}\mathbf{v}$.

Since all the subproblems of $\mathcal{P}5$ have closed-form optimal solutions, with $p_2$ iterations of AO, the total complexity is $O\Big(p_2[M^3_R+M^3_1+M^3_2+N^3_R+N^3_1+N^3_2]\Big)$.

\section{Simulation Results}
In this section, simulation results are provided to verify the proposed algorithm.
The number of transmit antennas is set as $(M_R,M_1,M_2)=(4,2,2)$ and the number of receive antennas is set as $(N_R,N_1,N_2)=(2,2,2)$.
All the channel entries are generated according to $\mathcal{CN}(0,\rho)$, with large scale fading $\rho=10^{-4}$.
It is assumed that the residual SI channels are further multiplied by SI coefficient $\kappa=0.1$ \cite{7}, and noise power $\sigma^2=-30\mathrm{dBm}$.
The same SINR QoS targets $\theta_1=\theta_2=\theta$ in dB are requested by all users.
Each point in the figures is obtained by averaging over 100 simulation runs, with independent channels in each run.

The following schemes are simulated:
the proposed FD AO scheme; the ideal FD scheme with no SI (solving $\mathcal{P}5$ without ZF constraints); the ZF-based FD AO scheme \cite{7}; the FD baseline scheme in Appendix B; the half-duplex AO scheme; and
the half-duplex baseline scheme.
For a fair comparison, the SINR target $\theta'$ for half-duplex schemes is $\theta'=(1+\theta)^2-1$ since the required data-rate for half-duplex schemes is $2\mathrm{log}(1+\theta)$.

We first analyze the total transmit powers of different schemes versus SINR target $\theta$.
As shown in Fig. 2a, the proposed FD AO scheme significantly outperforms the existing schemes, and approaches the ideal case very tightly over a wide range of SINR target $\theta$.
In particular, it saves $5\mathrm{dB}$ transmit power compared to the ZF-based FD AO scheme, which demonstrates the advantage of reserving a fraction of SI.
Besides, the FD schemes generally outperform the half-duplex schemes due to the reduced number of time slots, and are less sensitive to $\theta$.

\begin{figure}
  \centering
  \subfigure[]{
    \label{fig:subfig:b} %% label for second subfigure
    \includegraphics[width=3in]{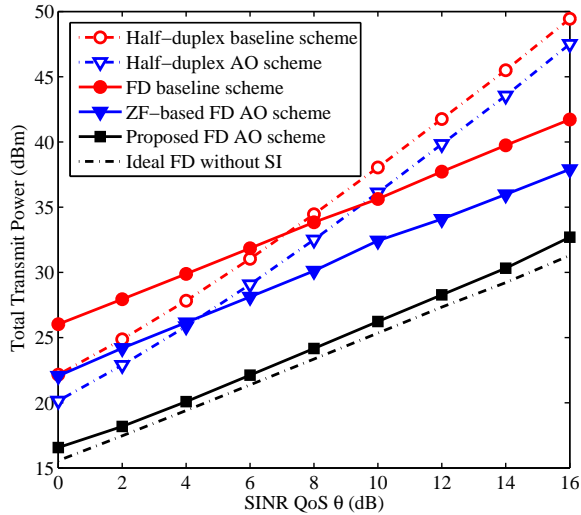}}
          \hspace{0.01in}
      \subfigure[]{
    \label{fig:subfig:b} %% label for second subfigure
    \includegraphics[width=3in]{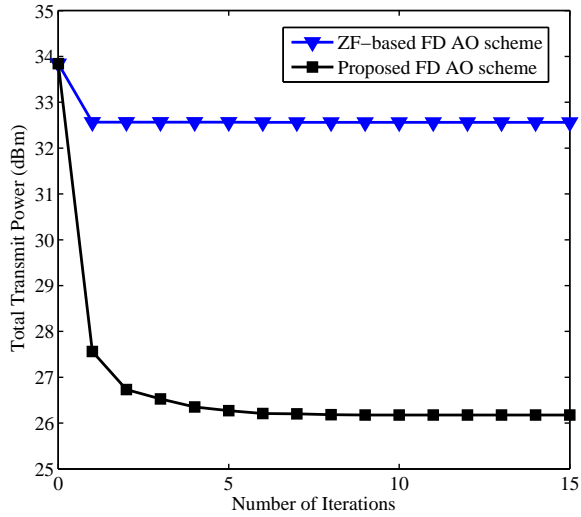}}
  \caption{Total transmit power for the case of $(M_R,M_1,M_2)=(4,2,2)$ and $(N_R,N_1,N_2)=(2,2,2)$ (a) Versus $\theta$ at noise power $\sigma^2=-30\mathrm{dBm}$; (b) Versus number of iterations at $\theta=10\mathrm{dB}$ and $\sigma^2=-30\mathrm{dBm}$.}
  \label{fig:subfig} %% label for entire figure
\end{figure}

To verify the convergence of the proposed FD AO algorithm, Fig. 2b shows the total transmit power versus number of iterations when $\theta=10\mathrm{dB}$.
It can be seen that while starting from the same initial point, the proposed FD AO algorithm shows much better performance than the ZF-based scheme after only $1$ iteration.
Furthermore, the number of iterations needed for the proposed FD AO algorithm to convergence is smaller than 10, which indicates fast convergence and moderate complexity of the proposed algorithm.

\section{Conclusion}
In this paper, we proposed a joint design of relay-user beamformers in FD-TWRC which reserves a fraction of SI.
The local optimum was obtained for relay beamformers, and the global optimum was obtained for user beamformers.
Furthermore, the alternating optimization algorithm was proved to be convergent.
Simulation results indicated that the proposed algorithm outperforms existing algorithms, and is close to the ideal case.

\appendices
\section{Proof of Property 1}
To prove part (i), consider the following inequalities
\begin{align}\label{property1-1}
&\Big|\Big(\mathbf{v}-\mathbf{v}^{[n]}\Big)^H\mathbf{g}_{R,i}\Big|^2\geq 0,~
\Big(\frac{\sqrt{\rho_i}}{\rho_i^{[n]}}-\frac{1}{\sqrt{\rho_i}}\Big)^2\geq 0.
\end{align}
Then from \eqref{property1-1}, we further have
\begin{align}
&\mathbf{v}^H\mathbf{\Phi}_i\mathbf{v}\geq
2\mathrm{Re}[(\mathbf{v}^{[n]})^H\mathbf{\Phi}_i\mathbf{v}]-(\mathbf{v}^{[n]})^H\mathbf{\Phi}_i\mathbf{v}^{[n]},
\nonumber\\
&
\frac{1}{\rho_i}\geq \frac{2}{\rho_i^{[n]}}-\frac{1}{(\rho_i^{[n]})^2}\rho_i,
\end{align}
which immediately proves part (i).
Part (ii) can be easily verified by substituting $\mathbf{v}^{[n]},\{\rho^{[n]}_{i}\}$ into the definition of $\widehat{\Upsilon}^{[n]}_{i},\widehat{\Delta}^{[n]}_{i}$.
To prove part (iii), we first calculate the following derivatives:
\begin{align}
&\partial\widehat{\Upsilon}^{[n]}_{i}/\partial\mathbf{v}=\Big[(\mathbf{v}^{[n]})^H\mathbf{\Phi}_i\Big]^T
,~\partial \Upsilon_{i}/\partial\mathbf{v}=
(\mathbf{v}^H\mathbf{\Phi}_i)^T
\\
&\partial\widehat{\Delta}^{[n]}_{i}/\partial\rho_{i}=
-\frac{1}{(\rho_i^{[n]})^2},~
\partial \Delta_{i}/\partial\rho_i=
-\frac{1}{\rho^2_i}.
\end{align}
Then by putting $\mathbf{v}=\mathbf{v}^{[n]},\{\rho_{i}=\rho^{[n]}_{i}\}$ into the above equations, the proof for part (iii) is completed.

\section{The Initialization of $\mathbf{v}^{[0]},\mathbf{w}^{[0]},\{\mathbf{f}^{[0]}_i,\mathbf{u}^{[0]}_i\}$}
To begin with,
set $\mathbf{w}^{[0]}=\sqrt{1/N_r}\mathbf{1}_{N_r}$ and $\mathbf{u}^{[0]}_i=\sqrt{1/N_i}\mathbf{1}_{N_i}$,
where $\mathbf{1}_{N}$ is an $N\times 1$ vector with all the elements being 1.
The initial $\mathbf{v}^{[0]},\{\mathbf{f}^{[0]}_i\}$ can be obtained from problem $\mathcal{P}5$.
Specifically, the feasibility condition for $\mathbf{f}_i$ in $\mathcal{P}5$ is given by
\begin{align}
|(\mathbf{w}^{[0]})^H\mathbf{H}_{3-i,R}\mathbf{f}_{3-i}|^2
>
\theta_i\sigma^2||\mathbf{w}^{[0]}||^2,~(\mathbf{u}^{[0]}_i)^H\mathbf{H}_{i,i}\mathbf{f}_i=0.
\nonumber
\end{align}
Using SVD $\mathbf{H}^H_{i,i}\mathbf{u}^{[0]}_i=[\mathbf{z}_{i}~\mathbf{Z}_{i}]\mathbf{\Lambda}_i\mathbf{V}_{i}^H$
and
$(\mathbf{w}^{[0]})^H\mathbf{H}_{i,R}\mathbf{Z}_{i}=\mathbf{A}_{i}\mathbf{B}_i[\mathbf{c}_{i}~\mathbf{C}_{i}]^H$
, we set
\begin{align}
\mathbf{f}^{[0]}_{i}
=\frac{\sqrt{2\theta_{3-i}\sigma^2||\mathbf{w}^{[0]}||^2}}{
|(\mathbf{w}^{[0]})^H\mathbf{H}_{i,R}\mathbf{Z}_{i}\mathbf{c}_{i}|}
\mathbf{Z}_{i}\mathbf{c}_{i}
.\nonumber
\end{align}

On the other hand, using SVD $\mathbf{H}^H_{R,R}\mathbf{w}^{[0]}=[\mathbf{z}_{R}~\mathbf{Z}_{R}]\mathbf{\Lambda}_R\mathbf{V}_{R}^H$
and
$\Big((\mathbf{u}_1^{[0]})^H\mathbf{H}_{R,1}+(\mathbf{u}_2^{[0]})^H\mathbf{H}_{R,2}\Big)\mathbf{Z}_{R}=\mathbf{A}_{R}\mathbf{B}_R[\mathbf{c}_{R}~\mathbf{C}_{R}]^H$
, the initial relay transmit beamformer is set as
$
\mathbf{v}^{[0]}=\sqrt{\alpha^{[0]}_R}\mathbf{Z}_{R}\mathbf{c}_{R}
,$
where
\begin{align}
\alpha^{[0]}_R=\mathop{\mathrm{max}}_{i=1,2}~
\frac{\theta_i\sigma^2\Big/|\mathbf{u}^H_i\mathbf{H}_{R,i}\mathbf{Z}_{R}\mathbf{c}_{R}|^2}{
|(\mathbf{w}^{[0]})^H\mathbf{H}_{3-i,R}\mathbf{f}^{[0]}_{3-i}|^2-\theta_i\sigma^2||\mathbf{w}^{[0]}||^2}. \nonumber
\end{align}

\end{document}